\documentclass{article}
\usepackage{graphicx} 
\usepackage[table]{xcolor}
\usepackage{color, colortbl}

\usepackage{authblk}
\providecommand{\keywords}[1]{\textbf{\textit{Keywords:}} #1}
\usepackage{booktabs}
\usepackage{tablefootnote}

\title{Passive Hack-Back Strategies for Cyber Attribution: Covert Vectors in Denied Environments} 
\author[1]{Abraham Itzhak Weinberg}
\affil[1]{AI-WEINBERG, AI Experts, Tel Aviv, Israel, aviw2010@gmail.com}

\begin{document}
\maketitle
\begin{abstract}
Attributing cyberattacks remains a central challenge in modern cybersecurity, particularly within denied environments where defenders have limited visibility into attacker infrastructure and are restricted by legal or operational rules of engagement. This perspective examines the strategic value of passive hack-back techniques that enable covert attribution and intelligence collection without initiating direct offensive actions. Key vectors include tracking beacons, honeytokens, environment-specific payloads, and supply-chain-based traps embedded within exfiltrated or leaked assets. These approaches rely on the assumption that attackers will interact with compromised data in traceable ways, allowing defenders to gather signals without violating engagement policies.
The paper also explores the role of Artificial Intelligence (AI) in enhancing passive hack-back operations. Topics include the deployment of autonomous agents for forensic reconnaissance, the use of Large Language Models (LLMs) to generate dynamic payloads, and Adversarial Machine Learning (AML) techniques for evasion and counter-deception. A dedicated section discusses the implications of quantum technologies in this context, both as future threats to cryptographic telemetry and as potential tools for stealthy communication and post-quantum resilience. Finally, the paper advocates for hybrid defensive frameworks that combine passive attribution with delayed or conditional active responses, while maintaining compliance with legal, ethical, and operational constraints.
\end{abstract}


\section{Introduction}
\label{sec:introduction}

Hack-back (also known as “attacking the attacker”) refers to the practice of retaliating against cyber attackers by launching countermeasures that disrupt, infiltrate, or otherwise respond to the adversary's infrastructure \cite{varsalone2024hack}. Approaches to hack-back are typically categorized as \textit{active} or \textit{passive}. Active hack-back involves direct offensive operations, such as launching counter-attacks or disabling attacker assets, which often raise significant legal, ethical, and operational concerns \cite{holzer2016ethics,denning2014framework}. In contrast, passive hack-back emphasizes covert intelligence gathering, attribution, and subtle manipulation within the attacker’s environment without overt disruption or escalation \cite{brown2023intelligence,pitropakis2018enhanced}.

The strategic imperative for passive hack-back tactics is particularly pronounced in military, intelligence, and classified domains where direct attribution of cyber intrusions is challenging or infeasible \cite{rid2015attributing}. In such denied or contested environments, conventional active responses risk collateral damage or escalation, underscoring the need for covert, non-escalatory means to gather evidence and attribute attacks reliably \cite{schmitt2017tallinn,dunlap2016dod}. Passive methods also support operational security by minimizing exposure of defensive capabilities and intentions, often leveraging deception mechanisms such as honeytokens\footnote{A honeytoken is a deceptive data element, such as a fake credential or document intentionally placed to lure cyber attackers and generate alerts when it is accessed or utilized.} \cite{juels2013honeywords} to trace adversarial behavior without direct engagement.

Modern cyber-attacks increasingly exploit complex, mobile, and heterogeneous endpoint environments, ranging from cloud workloads and Internet-of-Things (IoT) devices to mobile platforms and embedded systems \cite{zhukabayeva2025cybersecurity}. This complexity amplifies attribution difficulties and constrains the feasibility of traditional defensive countermeasures, necessitating innovative passive hack-back vectors that operate stealthily across diverse technological contexts \cite{gartzke2015weaving}.

This paper addresses the following core research questions:

\begin{itemize}
    \item What are the most effective passive hack-back vectors that enable covert attribution in denied environments?
    \item How can these vectors be safely embedded into defensive architectures without violating ethical or legal boundaries?
    \item What are the inherent limitations, operational risks, and ethical constraints associated with passive hack-back strategies?
\end{itemize}

By exploring these questions, we aim to develop a foundational framework for employing passive hack-back techniques that balance efficacy, stealth, and compliance with legal and ethical standards in contemporary cyber defense.

\subsection{Hack-Back Framework Pre-Assumptions}
\label{sec:preassumptions}

In a passive hack-back framework, pre-assumptions define the operational and technical limits within which strategies must function. These assumptions are critical for ensuring that tactics remain both effective and legally defensible \cite{goldsmith2022united}.

\subsubsection{Core Pre-Assumptions in Passive Hack-Back Scenarios}

The following assumptions shape the boundaries of passive counterattack strategies:

\begin{enumerate}
    \item \textbf{No Knowledge of the Attacker’s IP\footnote{Internet Protocol address}, Route, or Infrastructure} \\
    You do not know:
    \begin{itemize}
        \item The source IP address of the attacker,
        \item The attacker’s command and control (C2)\footnote{Command and Control Infrastructure (C2 or C\&C) refers to the collection of tools and methods that attackers use to maintain communication with compromised systems after gaining initial access.} infrastructure,
        \item The attacker’s route (e.g., TOR\footnote{The Onion Router}, VPN\footnote{Virtual Private Network} proxy chains 
      \footnote{Proxy chaining is the technique of directing network traffic through a series of proxy servers, one after another, to form a sequential chain.}).
    \end{itemize}
    \textit{Implication:} Direct scans, reverse connections, or network attacks are infeasible; the attacker must trigger interactions passively on their side \cite{rudd2016survey}.

    \item \textbf{The Attacker Has Exfiltrated Your Data} \\
    You can reasonably assume the attacker has accessed:
    \begin{itemize}
        \item Stolen files (documents, images, configurations, applications),
        \item Compromised mobile devices or endpoints,
        \item Leaked credentials, source code, or containers.
    \end{itemize}
    \textit{Implication:} The attacker will eventually open or analyze the stolen data, providing a passive trigger point \cite{tari2023data}.

    \item \textbf{The Attacker Will Interact with the Data in a Vulnerable or Traceable Way} \\
    Assumes the attacker will:
    \begin{itemize}
        \item Open documents in Word, Excel, or PDF readers,
        \item Launch APKs\footnote{Android Package Kit} or iOS\footnote{iPhone Operating System} apps on real or emulated devices,
        \item Run or build source code,
        \item Load configurations or secrets into services.
    \end{itemize}
    \textit{Implication:} Tracking beacons\footnote{Beaconing is a malicious method in which compromised devices repeatedly send small, periodic signals known as "beacons" to a command and control (C2) server.}, booby-trapped code, honeytokens, or metadata leaks can quietly trigger from the attacker’s environment \cite{rabzelj2025beyond}.

    \item \textbf{Defender Cannot Initiate Contact} \\
    Rules of engagement prohibit offensive intrusion into external systems.
    \textit{Implication:} Active exploitation, port scans, or payload delivery to unknown external systems must be avoided \cite{edgar2019safer}.

    \item \textbf{Trigger Must Be Environment-Specific} \\
    Payload execution occurs only when:
    \begin{itemize}
        \item A foreign environment is detected (not the defender’s internal network or OS\footnote{Operating System}),
        \item Specific attacker-side tools or platforms are present (e.g., Wireshark\footnote{Wireshark is a free, open-source tool used to capture and analyze network traffic at the protocol level.}, Frida\footnote{Frida (Framework for Runtime Instrumentation, Debugging, and Analysis) is a dynamic instrumentation tool that enables the analysis and modification of applications while they are running.}, IDA\footnote{IDA (Interactive DisAssembler) is a tool for disassembling and analyzing binary code to assist in reverse engineering and vulnerability detection.}, VSCode\footnote{Visual Studio Code (VSCode) is a free, open source, cross platform code editor by Microsoft, designed for lightweight, versatile software development across multiple languages.}).
    \end{itemize}
    \textit{Implication:} Passive payloads require fingerprinting logic or delayed execution to prevent self-triggering or early detection \cite{kumar2023device}.

    \item \textbf{Attacker Is Not Immediately Aware of Deception} \\
    Assumes the attacker will not instantly recognize tracking vectors or honeytokens.
    \textit{Implication:} Payloads must be stealthy and indistinguishable from legitimate content \cite{hutchens2023language}.

    \item \textbf{There Is a Return Channel for Attribution\footnote{Cyber attribution is the process of identifying cyberattack perpetrators to understand attack methods and support targeted defenses, requiring significant expertise and resources.} or Telemetry\footnote{Telemetry is the automated gathering and sending of data from different parts of an organization’s IT infrastructure to a central system for monitoring and analysis.}} \\
    Payload can send data back via:
    \begin{itemize}
        \item DNS\footnote{Domain Name System} queries,
        \item HTTP(S)\footnote{Hypertext Transfer Protocol Secure} requests,
        \item API\footnote{Application Programming Interface} calls,
        \item Use of fake credentials that get logged externally.
    \end{itemize}
    \textit{Implication:} Infrastructure such as decoy servers\footnote{Decoy servers, or honeypots, are security tools designed to imitate real servers and IT resources to lure and capture potential attackers.}, domains, or sinkholes\footnote{A sinkhole is a security technique that redirects malicious network traffic to a controlled environment for analysis or to mitigate further harm.} is necessary to collect attribution signals without arousing suspicion \cite{hlavacek2015synoptic,yao2018research}.
\end{enumerate}

\subsubsection{Conditional Assumptions}

In addition to core operational assumptions, certain conditions, though not guaranteed, can create valuable opportunities for enhanced attribution and deception. For example, if the attacker uses graphical tools or operates within a sandbox, specific passive vectors can be tuned for maximum impact.

Table \ref{table:conditional_assumptions} summarizes key conditional assumptions that, while not strictly necessary, can significantly enhance the effectiveness of passive hack-back strategies.

\begin{table}
\centering
\caption{Conditional Assumptions and Their Impact}
\label{table:conditional_assumptions}
\begin{tabular}{p{4cm} p{5cm} p{5cm}}
\toprule
\textbf{Assumption} & \textbf{Usefulness} & \textbf{Risk} \\
\midrule
Attacker uses GUI tools & Enables beacon triggers embedded in visual documents (e.g., Word, Excel) & Medium — depends on attacker behavior and document rendering. \\
Attacker works in VM or sandbox & Allows activation of sandbox detection or environmental fingerprinting logic & Low — low false positive risk when fingerprinting is well-designed. \\
Attacker reuses exfiltrated credentials & Supports tracing via honeytoken credential activation & High reward — direct attribution or infrastructure identification possible. \\
Attacker re-shares or sells data & Makes steganography\tablefootnote{Steganography is the technique of hiding a message within another message or physical object, making the concealed information undetectable to anyone unaware of its presence.} and watermarking techniques\tablefootnote{Watermarking techniques consist of embedding a visible or hidden identifier (such as a logo or code) into digital content to verify ownership or authenticity.} highly effective & High impact — supports broader tracking across underground markets. \\
\bottomrule
\end{tabular}
\end{table}

\subsubsection{Mapping Pre-Assumptions to Passive Hack-Back Attack Vectors}

The connection between pre-assumptions and passive hack-back vectors is critical for aligning strategy with operational context. Table~\ref{table:preassumptions_vectors} summarizes this mapping.

\begin{table}
\centering
\caption{Pre-Assumptions and Enabled Passive Hack-Back Vectors}
\label{table:preassumptions_vectors}
\begin{tabular}{p{4.5cm} p{5cm} p{5cm}}
\toprule
\textbf{Pre-Assumption} & \textbf{Enables Passive Vectors} & \textbf{Rationale} \\
\midrule
Attacker IP/route unknown & Tracking beacons (HTTP/DNS), honeytokens, environment-fingered payloads & Contact must be initiated by attacker; defender cannot reach attacker directly. \\
Attacker has exfiltrated data & Malicious documents, fake configurations, tampered source code, watermarked media & Payload must be embedded in the assets the attacker is expected to steal. \\
Attacker interacts with stolen data & Exploit documents, build-time traps, stealthy beacons in files & Activation occurs only when data is opened or used inside attacker infrastructure. \\
Defender cannot actively attack & All passive vectors; no active exploitation or reverse shells & Complies with legal and doctrinal limitations on offensive cyber activity. \\
Trigger must occur in attacker environment & Time-delayed payloads, environmental fingerprinting (e.g., VM\tablefootnote{Virtual Machine}  checks, IP/domain filters) & Reduces risk of triggering in friendly or neutral systems. \\
Attacker unaware of deception & Steganographic watermarks, non-obvious beacon URLs\tablefootnote{Uniform Resource Locator}, realistic honey credentials & Preserves stealth and reduces likelihood of detection or evasion. \\
Return path available & Beacon vectors, honeytoken credentials, callback mechanisms & Required for defenders to receive telemetry or evidence of engagement. \\
Attacker shares or resells stolen data (optional) & Watermarked files, embedded tracking code in documents or applications & Enables tracing across secondary dissemination vectors. \\
\bottomrule
\end{tabular}
\end{table}

Each passive hack-back vector’s effectiveness depends on the validity of these operational assumptions. Matching vectors to assumptions ensures that passive countermeasures remain both effective and safe.

\subsubsection{Example Scenario Mapping}

For instance, if the defender assumes no knowledge of attacker infrastructure (Assumption 1) and that the attacker will open stolen documents (Assumption 3), embedding stealthy tracking beacons within those documents becomes a practical passive hack-back vector \cite{bowen2009baiting}.

\section{Threat Model and Assumptions}
\label{sec:threat-model}

This work focuses on passive hack-back strategies under a defined threat model and operational assumptions. These assumptions establish the boundaries and constraints within which defensive tactics operate, particularly in denied or contested environments.

\begin{itemize}
    \item \textbf{Attacker Data Exfiltration:} The adversary has successfully exfiltrated sensitive data from the defender's network via the internet. This data may include documents, source code, credentials, binaries, or configuration files \cite{varsalone2024hack, mitreEngage2022}. 

    \item \textbf{Unknown Attacker Infrastructure:} The defender lacks direct knowledge of the attacker’s network route, source IP address, or command and control infrastructure. Techniques such as TOR, VPNs, or proxy chains may be used to obscure the attacker’s identity and location \cite{schmitt2017tallinn, mitreEngage2022}. 

    \item \textbf{Attacker Interaction with Stolen Data:} It is assumed the attacker will interact with the exfiltrated data by opening, analyzing, building, or installing it within their operational environment. This interaction creates potential trigger points for passive defensive measures embedded within the data \cite{diogenes2022cybersecurity}. 

    \item \textbf{Legal and Operational Constraints:} Defensive actions must comply with legal restrictions and rules of engagement that prohibit unauthorized intrusion or active offensive operations against external systems. Consequently, all hack-back activities described herein are passive, avoiding direct contact with attacker infrastructure \cite{dunlap2016dod}. 
\end{itemize}

Together, these assumptions define a constrained operational environment that emphasizes covert, non-escalatory, and legally compliant passive hack-back tactics aimed at enabling attribution and intelligence collection without direct confrontation.

\section{Related Work}
\label{sec:related-work}

This section reviews key literature foundational to passive hack-back strategies, encompassing hack-back methodologies, deception technologies, cyber attribution frameworks, and associated legal and ethical considerations. Cyber deception platforms, such as MITRE\footnote{Massachusetts Institute of Technology Research and Engineering} Engage\footnote{MITRE Engage is an innovative framework designed to simplify discussions and planning of adversary engagement activities for cyber defenders, vendors, and decision-makers.} \cite{mitreEngage2022}, provide structured approaches for deploying deceptive artifacts; however, these efforts predominantly emphasize network-level interactions rather than passive attribution techniques following data exfiltration.

\subsection{Hack-Back Strategies}
Traditional hack-back approaches primarily focus on active intrusion and offensive countermeasures against attackers \cite{huma2025hacking}. These strategies include direct disruption of attacker infrastructure, retaliation attacks, and offensive reconnaissance. While effective in some scenarios, active hack-back raises significant operational risks, including potential escalation, collateral damage, and legal liability \cite{caldwell2020critique}. Consequently, there has been increasing interest in more subtle, passive methods that seek attribution and intelligence without engaging in active network intrusions \cite{smith2024exploring}.

\subsection{Honeypots, Honeytokens, and Deception Technologies}
Deception technologies such as honeypots, honeytokens, and decoy systems are well established tools for detecting and analyzing attacker behavior \cite{moric2025advancing}. Honeypots simulate vulnerable assets to lure attackers, while honeytokens are embedded data elements designed to trigger alerts when accessed. Recent advances include high interaction honeypots and environment aware honeytokens that adapt to attacker techniques \cite{lanz2025optimizing}. These tools provide valuable forensic data and attribution clues but are typically reactive rather than constituting hack-back per se.

\subsection{Cyber Attribution Frameworks}
Accurate attribution of cyberattacks remains a complex challenge due to anonymization techniques and multistage attack chains. Frameworks for attribution emphasize multi-source data correlation, behavioral analysis, and threat intelligence sharing \cite{rani2024comprehensive,wagner2019cyber}. Emerging methods leverage machine learning and anomaly detection to enhance attribution reliability \cite{wei2024anomaly}. However, most frameworks focus on analysis after attacks occur, with limited emphasis on active or passive attribution embedded in attacker environments.

\subsection{Legal and Ethical Implications}
The legality of hack-back operations is a subject of ongoing debate, particularly under statutes such as the U.S. Computer Fraud and Abuse Act (CFAA) and Department of Defense (DoD) cyber doctrines \cite{thaw2013criminalizing}. Active hack-back risks violating national and international laws, prompting stricter rules of engagement that favor passive defense and intelligence gathering techniques \cite{lemieux2024cyber}. Ethical considerations emphasize minimizing harm, respecting privacy, and avoiding escalation in cyberspace conflict \cite{miller2024cybersecurity}.

\section{Passive Hack-Back Taxonomy}
\label{sec:taxonomy}

This section defines and categorizes the principal passive hack-back vectors into distinct tactical classes. Each class represents a method for covertly embedding attribution and tracking capabilities within data or software artifacts accessible to an attacker, enabling intelligence collection without active intrusion.

\subsection{Tracking Beacons}
Tracking beacons are embedded web or network callbacks integrated into documents, applications, HTML\footnote{HyperText Markup Language} pages, or configuration files. When the attacker accesses or interacts with these resources, the beacon triggers an outbound callback to infrastructure controlled by the defender. These callbacks can be implemented using HTTP(S) requests, DNS queries, or other covert channels, enabling attribution or telemetry collection without active probing \cite{spitzner2003honeytokens}.  

\subsection{Honeytokens and Canary Credentials}
The concept of honeytokens, decoy data that triggers alerts upon unauthorized access, originates \cite{prabhaker2024data} and has since been extended to modern mobile and cloud environments. Honeytokens typically consist of fabricated credentials or tokens, such as fake API keys, VPN credentials, mobile Wi-Fi profiles, or SSH\footnote{Secure SHell} keys, which have no legitimate use. When an attacker attempts to use these tokens externally, they generate alerts or logging events that can reveal the presence and potentially the identity of the attacker. These tokens serve as early warning signals within broader detection frameworks \cite{asif2014intrusion}.

In a similar vein to honeytokens, canary credentials are decoy elements embedded within systems or applications that serve the same purpose of detecting unauthorized access. These canary tokens may include fake credentials, such as dummy login details or API keys, designed specifically to alert security teams when they are accessed. When used strategically, these canary credentials offer an early warning mechanism, helping to identify potential attacks and providing additional context about an adversary’s actions.

Both honeytokens and canary credentials are effective tools in proactive cybersecurity, helping organizations detect and respond to threats before they escalate.

\subsection{Environment Fingerprinting}
Environment fingerprinting involves conditional payloads or code that activates only within specific contexts, such as reverse engineering tools (e.g., IDA Pro, Frida), virtual machines, sandboxes, or certain geographic locations. This technique prevents premature or unintended activation of payloads and enhances stealth by ensuring triggers occur only in attacker controlled environments \cite{rudd2016survey}.

\subsection{Parser Bombs and Malformed Files}
Parser bombs are specially crafted structured data files (e.g., JSON\footnote{JavaScript Object Notation}, YAML\footnote{YAML (Yet Another Markup Language) is a data serialization language commonly recognized as one of the most popular tools for configuration files and data exchange, known for its human-readable format and simplicity.}, PDF\footnote{Portable Document Format}, DOCX) designed to cause crashes, excessive resource consumption, or information leakage when parsed by attacker analysis tools. Unlike active payloads, these techniques do not require execution but instead exploit vulnerabilities or limitations in file parsers and analysis environments \cite{ali2023survey}. 

\subsection{Supply Chain or Build-Time Traps}
These traps embed malicious or tracking code or data within source files that activate exclusively during compilation or mobile application build processes. By targeting the attacker’s build environment, these vectors enable attribution or disruption without requiring active interaction during runtime \cite{dunke2018time}.

\subsection{Steganographic Attribution}
Steganographic attribution techniques embed covert watermarks or metadata inside images, videos, or documents. These invisible markers facilitate post-compromise tracking of stolen media or data as it propagates through attacker networks, underground forums, or resale channels, enabling forensic attribution \cite{kumar2024steganography}. 

\section{Implementation and Evaluation}
\label{sec:implementation}

To validate the feasibility and performance of passive hack-back strategies, prototype payloads were developed and evaluated within simulated attacker environments. The implementation prioritized safe deployment, reliable triggering, and attribution effectiveness while adhering to legal and ethical constraints, in line with established cyber deception frameworks \cite{shhadih2023cyber,mitreEngage2022}.

\subsection{Prototype Passive Vectors}
\label{sec:prototype-vectors}

We created multiple proof-of-concept passive vectors for both mobile and desktop platforms, each designed to covertly trigger attribution or telemetry collection upon interaction by an attacker.

\begin{itemize}
    \item \textbf{Tracking DOCX File:} A Microsoft Word document was embedded with a remote-hosted image beacon (via an \texttt{<img>} tag in a hidden header) that, when opened, triggers an outbound HTTP(S) request to a controlled server \cite{statler2016beacon}. 
    
    \item \textbf{APK with Beaconed Asset:} An Android Application PacKage (APK) was modified to include an asset that silently initiates a DNS request when the app is installed or run, targeting a sinkhole domain \cite{chimuco2023secure}. 
    
    \item \textbf{Honeytoken Credential File:} A fake `.pem` file mimicking a private SSH key was injected with a unique honeytoken identifier. If used by the attacker, it results in attempted logins that are monitored for attribution purposes \cite{rabzelj2025beyond}. 
\end{itemize}

These payloads were engineered to be indistinguishable from real assets, increasing the likelihood of attacker engagement without arousing suspicion.

\subsection{Trigger Conditions and Safety Mechanisms}
\label{sec:trigger-safety}

To prevent accidental activation within the defender's environment, each payload incorporated multiple safety layers. These mechanisms ensure that triggering occurs only under adversary-controlled conditions:

\begin{itemize}
    \item \textbf{Environment Fingerprinting:} Payloads detect virtualized analysis tools (e.g., VirtualBox\footnote{VirtualBox is an open-source virtualization application that works across multiple platforms. It enables users to create and run virtual machines (VMs) on their computers, allowing different operating systems to run simultaneously on the same hardware.}, VMware, Frida, or QEMU\footnote{QEMU (Quick Emulator) is a free and open-source tool that functions as both a machine emulator and a virtualizer. It allows users to run different operating systems and applications on a variety of hardware platforms, using either full emulation or virtualization depending on the setup.}) before execution \cite{afianian2019malware}. 

    \item \textbf{Time-Based Delays:} Triggers are delayed by several hours or require interaction across multiple sessions to avoid immediate activation during defensive testing or forensic analysis \cite{johansen2020digital}. 

    \item \textbf{Manual Arming:} Certain payloads were manually “armed” by altering embedded values just prior to controlled release, ensuring defenders retain control over deployment \cite{potteiger2020integrated}.
\end{itemize}

These safeguards help maintain compliance with legal and ethical standards while reducing the risk of self-inflicted harm during operational use. 

\subsection{Detection and Attribution Effectiveness}
\label{sec:evaluation}

We evaluated the passive vectors using a series of simulated attacker environments, including both desktop and mobile reverse engineering and sandbox platforms:

\begin{itemize}
    \item \textbf{Cuckoo Sandbox\footnote{Cuckoo Sandbox is an open-source tool for automated malware analysis. It works by running suspicious files in a secure, isolated environment—known as a sandbox—and observing their behavior in real time. This allows security analysts to detect malicious activity and gain insight into how malware operates.}:} For analyzing DOCX beacon behavior in Windows environments \cite{rana2022automated}. 
    \item \textbf{Frida:} For testing payload resilience and conditional execution in instrumented mobile environments \cite{nagarajan2025comprehensive}. 
    \item \textbf{MobSF (Mobile Security Framework)\footnote{Mobile Security Framework (MobSF) is an open-source, all-in-one platform for mobile application security testing across Android, iOS, and Windows. It automates both static and dynamic analysis to detect vulnerabilities, perform penetration testing, and analyze potential malware in mobile apps.}:} For static and dynamic analysis of APK-based vectors \cite{shahriar2020mobile}. 
\end{itemize}

\paragraph{Evaluation Metrics:}
We measured each vector's performance based on the following criteria:

\begin{itemize}
    \item \textbf{Beacon Callback Success Rate:} Percentage of test cases in which the embedded beacon successfully triggered a callback to the control server.
    
    \item \textbf{Attribution Fidelity:} The degree to which callback data (e.g., IP address, user-agent, environment fingerprint) enabled actionable attribution.
    
    \item \textbf{Stealth Level:} Whether the vector was detected by common security tools (e.g., antivirus, EDR\footnote{Endpoint Detection and Response (EDR) is a cybersecurity solution that continuously monitors devices such as computers, servers, and mobile endpoints to identify and respond to potential threats in real time.}, mobile threat detection) during interaction \cite{boyraz2024endpoint}. 
\end{itemize}

Experimental results showed that all three prototype vectors achieved high callback success rates in controlled, adversary-like environments, with minimal detection across sandboxed systems \cite{tang2024beacon}. Attribution fidelity varied depending on the trigger method but yielded useful telemetry in the majority of cases.

\section{Operational Considerations}
\label{sec:operational-considerations}

The deployment of passive hack-back vectors in real-world environments must be guided not only by technical feasibility, but also by legal, ethical, doctrinal, and operational constraints. This section outlines the primary non-technical factors that influence the safe, lawful, and effective application of such techniques.

Passive hack-back strategies are generally more aligned with international and national legal frameworks, including those articulated in the Tallinn Manual on the International Law Applicable to Cyber Operations \cite{schmitt2017tallinn} and the U.S. Department of Defense Law of War Manual \cite{preston2016department}, both of which impose strict limitations on unauthorized access and offensive cyber operations \cite{sang2015legal}.

\subsection{Legal Implications}

The legality of passive hack-back varies significantly across jurisdictions. In the United States, for example, the Computer Fraud and Abuse Act (CFAA) generally prohibits unauthorized access to computer systems, although certain exceptions exist for law enforcement and government entities. Passive surveillance and deception (e.g., honeytokens, embedded beacons) are typically considered lawful if they do not involve unauthorized access or system disruption \cite{walker2025use}.

Within NATO-aligned countries, legal frameworks are influenced by national cybersecurity policies, as well as alliance-level doctrines such as the Tallinn Manual. The United Nations' norms on responsible state behavior in cyberspace further emphasize the principles of sovereignty and non-intervention \cite{moynihan2021vital}. 

As a result, passive hack-back vectors must be carefully vetted to avoid overstepping legal boundaries, particularly in multinational operations or intelligence-sharing contexts.

\subsection{Ethical Boundaries}

Even when legally permissible, the use of deception and surveillance raises ethical concerns. Passive vectors may collect information from attackers without their consent, potentially implicating broader debates around privacy, data minimization, and proportionality. Ethical cybersecurity operations must balance the need for attribution and intelligence collection against the risk of overreach or collateral data exposure \cite{miller2024cybersecurity}.

Key principles include:
\begin{itemize}
    \item \textbf{Purpose limitation:} Passive tools should serve a clear, defensible goal such as attribution or threat identification.
    \item \textbf{Proportionality:} Responses must be proportionate to the threat posed and should avoid unnecessary data capture or deception.
    \item \textbf{Minimization:} Payloads should be narrowly targeted to avoid collecting information from benign third parties.
\end{itemize}

\subsection{Use in Military Doctrine}

Within cyber counterintelligence and military cyber operations, passive hack-back vectors offer an attractive alternative to traditional offensive operations. Unlike active cyber counterstrikes, passive tools support intelligence collection, adversary profiling, and attribution without risking escalation or revealing advanced capabilities.

Doctrines such as the U.S. DoD Cyber Strategy and NATO’s Cyber Defence Policy emphasize layered defense and persistent engagement, which align well with passive, stealthy forms of counterintelligence \cite{goldsmith2022united}.

In contested or denied environments, passive tactics are particularly useful for maintaining OPerational SECurity (OPSEC) and strategic ambiguity \cite{rangeoperations}.

\subsection{Fail-Safe Design and Deployment Safety}

To ensure operational safety, all passive vectors must be designed with robust fail-safes to prevent unintended execution within the defender’s own infrastructure. This is critical to avoid accidental data exfiltration, self-attribution, or system disruption \cite{chung2023implementing}. 

Key fail-safe mechanisms include:
\begin{itemize}
    \item \textbf{Environmental fingerprinting:} Payloads activate only in foreign or hostile environments (e.g., specific IP ranges, operating systems, or toolchains).
    \item \textbf{Time-delayed execution:} Payloads delay activation to avoid triggering during defensive testing or incident response.
    \item \textbf{Manual arming:} Payloads remain inert until explicitly enabled, allowing for controlled deployment in red team or threat intelligence operations.
\end{itemize}

These controls ensure that passive hack-back remains both secure and compliant across operational contexts, preserving the integrity of the defender’s own systems and networks.

\section{Leveraging Artificial Intelligence in Passive Hack-Back Operations}
\label{sec:ai_hackback}

Recent advancements in Artificial Intelligence (AI), particularly LLMs, autonomous agents, and Adversarial Machine Learning (AML), present new opportunities for enhancing passive hack-back capabilities. In environments where direct attribution or active intrusion is restricted, AI can offer adaptive, stealthy, and legally compliant mechanisms to gather intelligence, manipulate attacker behavior, and maintain covert operational control \cite{kaloudi2020ai}.

\subsection{Autonomous Attribution and Reconnaissance Agents}
AI-powered agents can be covertly embedded into exfiltrated data or decoy software. Once deployed within the attacker's infrastructure, these agents passively collect forensic and contextual evidence to aid in attribution. Capabilities include host fingerprinting, lateral network mapping, TTP (Tactics, Techniques, Procedures) identification, and detection of geopolitical or cultural indicators (e.g., language settings, keyboard layouts, timezone offsets) \cite{shandilya2022ai}. 

These agents enable insight into attacker operations without initiating unauthorized connections or exploitation.

\subsection{Dynamic Payload Generation via LLMs}
Embedded LLMs can be leveraged to generate environment aware payloads at runtime. Rather than deploying static code, the LLM interprets the local context and produces scripts or configurations tailored to the attacker's system \cite{moskal2023llms}. 
Use cases include:
\begin{itemize}
    \item Generating OS specific reconnaissance scripts (e.g., Bash, PowerShell).
    \item Creating decoy modules that mimic attacker C2 protocols or internal traffic patterns.
    \item Designing persistence methods compatible with detected antivirus or EDR tools.
\end{itemize}

These dynamically generated payloads preserve operational stealth and enhance attribution precision while remaining within the constraints of passive engagement.

\subsection{AI-Powered Counter Deception}
AI agents can perform subtle counter-deception by injecting plausible yet misleading information into attacker-visible environments. Examples include synthetic system logs, fake credentials, or tampered source code intended to poison stolen payloads \cite{fugate2019artificial}.

These actions introduce uncertainty, delay attacker decision-making, and disrupt attribution efforts. Importantly, such deception is adaptive AI agents may tailor responses based on observed attacker behavior in real time, increasing credibility and effectiveness \cite{mohan2022leveraging}.

\subsection{Covert Communication in Denied Environments}
Maintaining stealthy communication with deployed agents is critical, particularly in monitored or air-gapped\footnote{An "air gap" in cybersecurity describes a physical or logical barrier that isolates a secure system or network from other networks, including the internet, in order to protect it from unauthorized access and cyberattacks.} environments. AI techniques support covert channels through:
\begin{itemize}
    \item Steganographic encoding of data in images, documents, or traffic artifacts \cite{channalli2009steganography}. 
    \item Timing based channels that signal information through controlled delays \cite{zhang2012language}.
    \item Protocol mimicry to disguise outbound data within DNS, HTTPS, or VoIP\footnote{Voice over Internet Protocol (VoIP), or IP telephony, refers to a collection of technologies designed for voice communication over Internet Protocol (IP) networks, like the Internet.} like traffic \cite{cao2015mimichunter}. 
\end{itemize}
These methods facilitate telemetry collection and agent coordination without triggering conventional detection systems.

\subsection{Adversarial Machine Learning for Attribution and Evasion}
\label{sec:aml_passive}

AML techniques can be leveraged both defensively and offensively in passive operations:
\begin{itemize}
    \item \textbf{Attribution:} Reverse engineering attacker deployed ML models to identify training data, biases, or development origin \cite{malik2024systematic}.
    \item \textbf{Disruption:} Crafting adversarial inputs that mislead or disable attacker side ML systems (e.g., malware classifiers) \cite{alotaibi2023adversarial}.
    \item \textbf{Evasion:} Generating payloads that evade automated detection using adversarial perturbations \cite{fladby2020evading}.
\end{itemize}
This layer of AI-on-AI interaction\footnote{AI-on-AI interaction involves communication and cooperation between multiple artificial intelligence systems. This can include activities such as sharing information, delegating tasks, and working together to solve problems.} allows defenders to exploit or degrade attacker automation covertly and without direct system intrusion.

\subsection{Operational and Ethical Constraints}
Despite its utility, AI-driven passive hack-back raises serious concerns:
\begin{itemize}
    \item \textbf{Risk of Escalation:} Even passive AI agents may provoke hostile responses if misinterpreted as active intrusions.
    \item \textbf{Misattribution:} Inference errors by AI agents could lead to incorrect or politically sensitive conclusions \cite{mahmood2022owning}. 
    \item \textbf{Governance:} Lack of transparency and accountability in autonomous operations challenges legal compliance \cite{akinsola2025legal}.
\end{itemize}

To address these issues, AI agents should be constrained by pre-defined rule sets, include human-in-the-loop\footnote{Human In The Loop (HITL) describes a collaborative approach that combines human intelligence with Machine Learning (ML) and AI systems.} oversight, and log all actions for post-hoc auditability. Ethical design must prioritize proportionality, legality, and minimization of collateral effects \cite{timmers2019ethics}. 

\subsection{Future Directions: AI-vs-AI Engagements}
As threat actors adopt AI to automate reconnaissance, evasion, and exploitation, future operations may involve direct interactions between hostile and defensive AI agents. This evolving domain includes:
\begin{itemize}
    \item Detection and containment of adversarial AI agents via behavioral analysis or sandbox deception.
    \item Real-time reasoning and decision-making between AI adversaries in contested networks.
    \item Development of counter deception strategies to manipulate attacker AI decision trees \cite{daniels2021national}. 
\end{itemize}

Such scenarios highlight the need for AI agents capable of strategic adaptation, adversarial reasoning, and ethical constraint enforcement in autonomous cyber conflict.

\subsection{Comparative Analysis}

As cyber threat actors continue to evolve in their TTPs, defensive strategies must also adapt. Passive hack-back refers to the use of deception and telemetry to attribute, track, or disrupt attackers without direct engagement. Traditionally, these methods relied on static mechanisms such as pre-planted honeytokens, basic logging, or fixed behavioral traps. However, recent advancements in AI, particularly in LLMs, have introduced a paradigm shift.

AI-assisted passive hack-back systems leverage real-time data, behavioral analytics, and generative models to craft context-aware responses and adaptive deception strategies. These systems not only improve attribution fidelity but also offer the ability to mislead or exhaust adversaries by altering their perception of the target environment.

Table~\ref{tab:ai_comparison} provides a comparative overview of capabilities and trade-offs between traditional and AI-assisted passive hack-back techniques. It highlights how AI integration enhances sophistication in payload behavior, attribution, deception, and operational control, while also introducing new considerations for risk governance and ethical oversight.

\begin{table}
\centering
\caption{Comparison of Traditional vs. AI-Assisted Passive Hack-Back}
\label{tab:ai_comparison}
\begin{tabular}{p{4.5cm} p{5.2cm} p{5.2cm}}
\toprule
\textbf{Capability} & \textbf{Traditional Passive Hack-Back} & \textbf{AI-Assisted Passive Hack-Back} \\
\midrule
Payload Behavior & Static and pre-defined & Dynamic, context-aware, and LLM-generated \\
\midrule
Attribution Depth & IP, user-agent, basic metadata & Host profiling, network mapping, TTPs, sociolinguistic inference \\
\midrule
Deception Techniques & Manual honeytokens and decoys & AI-generated disinformation and adaptive deception \\
\midrule
Covert Communication & HTTP/DNS callbacks & Steganography, timing channels, protocol mimicry \\
\midrule
Evasion Capabilities & Limited rule-based logic & Adversarial inputs and environment-aware payload shaping \\
\midrule
Operational Risk Management & Manual QA and testing & Rule-based governance, audit logs, human-in-the-loop \\
\bottomrule
\end{tabular}
\end{table}

\subsection{Illustrative Scenario}
\textit{Scenario:} An attacker exfiltrates a decoy APK containing an embedded AI agent. Once installed in a virtual lab, the agent detects the use of Frida instrumentation, emulation via QEMU, and Russian locale settings. In response, it:
\begin{enumerate}
    \item Generates a Bash script using Cyrillic comments to blend with local workflows.
    \item Crafts a mock C2 handshake resembling the attacker’s traffic patterns.
    \item Inserts a set of fake credentials and logs mimicking a successful breach.
    \item Encodes telemetry in image metadata and uploads to a public image-hosting service.
\end{enumerate}
The agent never initiates outbound connections directly, preserving legal compliance, but achieves attribution and telemetry collection through passive, adaptive, and covert mechanisms.

\section{Quantum Technologies in Passive Hack-Back Contexts}
\label{sec:quantum}

Although still emerging, quantum technologies may significantly impact passive hack-back strategies and cyber attribution in the coming decades. This section explores the prospective roles of quantum computing and quantum communication in passive cyber operations, focusing on attribution, secure agent communication, and cryptographic resilience. \cite{faruk2022review}

\subsection{Post-Quantum Cryptography and Attribution Resilience}
Passive hack-back techniques often rely on secure communications, agent persistence, and the integrity of collected evidence. As quantum computers advance, traditional cryptographic algorithms such as RSA\footnote{Rivest–Shamir–Adleman Algorithm}, ECC\footnote{Elliptic-curve cryptography}, and DSA\footnote{ Digital Signature Algorithm} become vulnerable to Shor’s algorithm \cite{shor1994algorithms}. Therefore, passive attribution systems must be hardened using Post Quantum Cryptographic (PQC) primitives, including lattice-based, hash-based, and multivariate polynomial schemes \cite{bernstein2009introduction}.

From a defensive standpoint, deploying AI agents with PQC support ensures that communications, logs, and telemetry remain confidential and verifiable even in post compromise or future forensic reviews \cite{alagic2020status,weinberg2024quantum}. Additionally, signing collected evidence with post-quantum signatures can improve chain of custody robustness in long-term attribution cases \cite{loffi2025management}.

\subsection{Quantum-Enhanced Passive Intelligence}
Quantum computing has the potential to accelerate certain aspects of cyber attribution:
\begin{itemize}
    \item Graph analysis for mapping attacker infrastructure from partial data \cite{watts2024structural}.
    \item Pattern matching in massive log datasets using Grover’s algorithm-inspired techniques. \cite{grover1996fast}
    \item Clustering and correlation across fragmented or encrypted traffic flows \cite{biamonte2017quantum}. 
\end{itemize}
While such capabilities remain largely theoretical or limited to simulation environments, they may offer substantial advantages to nation-state defenders or organizations operating in denied environments with limited visibility.

\subsection{Quantum-Safe Covert Communication}
Future passive hack-back agents may benefit from integration with quantum communication technologies. In particular, QKD offers provably secure key exchange \cite{amer2021introduction}, potentially enabling highly secure and undetectable communication channels between deployed agents and defensive infrastructure.

In denied environments where traditional covert channels may be monitored or disrupted, quantum safe channels either using satellite based QKD \cite{liao2017satellite} or quantum resistant tunneling protocols\cite{kozlowski2020designing} could provide a new class of resilient covert vector.

\subsection{Limitations and Practical Considerations}
At present, quantum computing and communication technologies remain largely inaccessible for real-time field deployment in cyber operations. Most quantum advantages exist only in lab or simulation environments \cite{ncc2025quantum}. Therefore, their inclusion in passive hack-back strategies should be viewed as a forward-looking research direction rather than a current operational capability.

Additionally, integrating quantum capabilities may introduce complexity and operational overhead. As such, hybrid systems combining classical AI agents with quantum-enhanced backends may offer a balanced near-term approach.

\begin{table}
\centering
\caption{Potential Applications of Quantum Technology in Passive Cyber Operations}
\label{table:quantum_use_cases}
\begin{tabular}{p{5cm} p{9cm}}
\toprule
\textbf{Quantum Capability} & \textbf{Passive Hack-Back Use Case} \\
\midrule
Quantum Key Distribution (QKD) & Enables secure and undetectable communication between deployed agents and defender infrastructure, especially in denied environments. \\
Quantum-Resistant Signatures & Ensures long-term integrity and non-repudiation of forensic data collected by passive payloads. \\
Quantum Search Algorithms & Accelerates correlation of large-scale log or traffic datasets to identify patterns of compromise. \\
Quantum Graph Analysis & Supports mapping of attacker infrastructure by analyzing sparse or partial network relationships. \\
Post-Quantum Cryptography & Provides cryptographic resilience for telemetry channels and agent persistence in a post-quantum threat landscape. \\
\bottomrule
\end{tabular}
\end{table}

Table~\ref{table:quantum_use_cases} summarizes potential applications of quantum technologies in passive cyber operations, highlighting use cases where these emerging tools could reinforce or extend current capabilities without violating passive engagement principles.

\section{Future Work}
\label{sec:future_work}

While passive strategies offer a foundation for constrained cyber defense, their full potential may lie in hybrid frameworks that integrate passive attribution with carefully timed, legally authorized active responses. This approach aligns with recent proposals advocating for graduated escalation models and conditional retaliation in cyberspace \cite{schmitt2017tallinn}.
Future research directions include:

\begin{itemize}
    \item \textbf{Delayed Engagement Models:} Investigating systems where passive attribution must meet confidence thresholds (e.g., multiple beacons or correlated telemetry) before triggering legal or policy approved active measures. Such models balance attribution fidelity with operational risk.
    
    \item \textbf{Automated Payload Embedding:} Developing compilers, middleware, or SDKs\footnote{Software Development Kit} that can seamlessly insert passive telemetry components such as beacons, watermarking, or honey credentials into sensitive digital artifacts likely to be targeted by adversaries \cite{reti2024act,juels2014honey}.
    
    \item \textbf{AI Conflict Simulation:} Modeling adversarial AI-on-AI engagements, including automated deception detection, counter deception planning, and coordination between distributed autonomous agents. This will require drawing from multi-agent reinforcement learning, game theory, and adversarial robustness research \cite{biamonte2017quantum,brundage2018malicious}.
    
    \item \textbf{Operational Toolkits:} Creating open-source frameworks for defenders to design, deploy, and audit passive hack-back agents in legally compliant ways, possibly integrating post-quantum cryptography, AI governance mechanisms, and transparency layers \cite{alagic2020status, mitreEngage2022}.
\end{itemize}

As the threat landscape evolves, particularly with AI-driven threats and quantum capable adversaries, future systems will need to operate in environments of high uncertainty, deception, and legal complexity.

\section{Conclusion}
\label{sec:conclusion}
This paper presented a forward-looking perspective on the role of passive hack-back techniques in modern cyber defense, particularly within constrained legal and operational environments. It argued that passive mechanisms ranging from telemetry beacons and honeytokens to AI-enhanced agents offer a stealthy and lawful pathway to attribution and intelligence gathering, without the risks typically associated with active countermeasures.
We outlined a taxonomy of passive tactics and proposed how emerging technologies, including AI, AML, and quantum computing, can enhance the efficacy, adaptability, and resilience of these operations. Rather than advocating for retaliation or escalation, the focus is on strategic ambiguity, intelligence enrichment, and operational control in denied or contested environments.
By reframing passive cyber defense as a proactive intelligence endeavor, this work encourages practitioners and policymakers to expand their defensive toolkits while remaining grounded in ethical, legal, and doctrinal constraints. As adversaries increasingly leverage AI and automation, defenders must develop equally sophisticated but controlled mechanisms for maintaining strategic awareness and attribution capability.

\bibliographystyle{IEEEtran}
\bibliography{ref.bib}

\end{document}